\documentclass[aps,pra,twocolumn,superscriptaddress,longbibliography,10pt]{revtex4-1}
\usepackage{times}

\usepackage[usenames,dvipsnames]{xcolor}

\usepackage{tikz}
\usetikzlibrary{arrows,shapes,positioning,shadows,backgrounds,fit}

\usepackage[english]{babel}
\usepackage{graphicx, dcolumn, bm, color, amsmath, relsize, amsmath, dsfont, mathrsfs, empheq, verbatim, upgreek, etoolbox}
\usepackage[caption = false]{subfig}

\usepackage[colorlinks=true,linkcolor=blue,urlcolor=blue,citecolor=blue]{hyperref}
\usepackage{hyperref}
\usepackage{graphics}
\usepackage{bm}
\usepackage{color}
\usepackage{amscd}
\usepackage{amsfonts}
\usepackage{amssymb}
\usepackage{graphicx}
\usepackage{tabularx}

\usepackage{dcolumn}
\usepackage{bm}
\usepackage{graphicx}
\usepackage{amsmath}
\usepackage{latexsym}
\usepackage{array}
\usepackage{epsfig}

\newcommand{\Ham}{\mathcal{H}}
\newcommand{\Id}{\mathds{1}}

\newcommand{\ket}[1]{\left\vert#1\right\rangle}
\newcommand{\bra}[1]{\left\langle#1\right\vert}

\begin{document}

\newdimen\origiwspc
\newdimen\origiwstr

\title{Non-equilibrium steady-states of memoryless quantum collision models}

\author{Giacomo Guarnieri}
\email{guarnieg@tcd.ie}
\affiliation{School of Physics, Trinity College Dublin, College Green, Dublin 2, Ireland}

\author{Daniele Morrone}
\affiliation{Dip. Fisica, Universit\'{a} della Calabria, 87036 Arcavacata di Rende (CS), Italy}

\author{Bar\i\c{s} \c{C}akmak}
\affiliation{College of Engineering and Natural Sciences, Bah\c{c}e\c{s}ehir University, Be\c{s}ikta\c{s}, Istanbul 34353, Turkey}

\author{Francesco Plastina}
\affiliation{Dip. Fisica, Universit\'{a} della Calabria, 87036 Arcavacata di Rende (CS), Italy}
\affiliation{INFN-Gruppo Collegato di Cosenza, 87036, Cosenza, Italy}

\author{Steve Campbell}
\affiliation{School of Physics, University College Dublin, Belfield, Dublin 4, Ireland}

\date{\today}

\begin{abstract}
We investigate the steady state properties arising from the open system dynamics described by a memoryless (Markovian) quantum collision model, corresponding to a master equation in the ultra-strong coupling regime. By carefully assessing the work cost of switching on and off the system-environment interaction, we show that only a coupling Hamiltonian in the energy-preserving form drives the system to thermal equilibrium, while any other interaction leads to non-equilibrium steady states that are supported by steady-state currents. These currents provide a neat exemplification of the housekeeping work and heat. Furthermore, we characterize the specific form of system-environment interaction that drives the system to a steady-state exhibiting coherence in the energy eigenbasis, thus, giving rise to families of states that are non-passive.
\end{abstract}

\maketitle

\section{Introduction}
Quantum thermodynamics~\cite{DeffnerCampbellBook,JPA_Goold,arXiv_Alicki,Janet2016} aims at describing energy exchanges, in the form of work and heat, between quantum systems, and allows for the study of the quantum source of irreversibility in non-equilibrium processes. The comprehension of such energetic exchanges is vital to extend our understanding of the laws of thermodynamics in the quantum regime, while also paving the way to exploit the peculiarities of quantum systems to realize efficient next generation quantum technologies.

While exquisite experimental control of many types of quantum systems and devices has been achieved to date, the unavoidable interaction with their surroundings has detrimental effects on their genuinely quantum properties, notably coherence and quantum correlations. The framework of open quantum systems allows us to account for these environmental effects through methods and models that provide a very accurate description of the system's evolution~\cite{BreuerPetruccione,AlickiLendiBook}. A particularly useful approach is given by quantum collision models, which constitute a very rich platform to simulate open quantum systems in a simple and yet reliable  way~\cite{RauPR, Scarani2002, BuzekPRA, ziman2005b, ziman2010, ciccarello2017}. Owing to their high flexibility, collision models have been utilized as a tool to study various physical phenomena, such as memory effects (or quantum non-Markovianity)~\cite{RuariPRA, StrunzPRA2016, CampbellPRA2018, CampbellNJP2019, BarisPRA2017}, information erasure~\cite{RuariPRL, PezzuttoNJP}, quantum thermodynamic processes~\cite{BarraSciRep, GabrieleNJP2018, Obinna2019, OzgurDiode, FilippovPRA}, quantum synchronization~\cite{BarisPRA2019}, and the quantum-to-classical transition~\cite{CampbellPRA2019, garcia2019decoherence}.

In this work, we exploit the versatility of collision models to fully characterize the steady-state thermodynamics for arbitrary system-environment interactions in the fast-collision time limit, complementing the study of equilibration and thermalization via collision models performed in Refs.~\cite{BarraPRE2017, SeahPRE2019, Onat2019, Parrondo}. In particular, we consider a system that interacts with an environment, consisting of an array of identical thermal constituents, in a memoryless (Markovian) collision model setting. We show that, while the dynamics is fully described by a Gorini-Kossakowski-Sudarshan-Lindblad (GKSL) master equation, care must be taken in assessing the thermodynamics of the interaction process. More specifically, work, heat, and entropy production cannot generally be derived solely from the system's degrees of freedom, but one needs to account for the interaction energy as well. Once this is satisfactorily taken into account and fully exploited, one is led to an elegant demonstration of the so-called housekeeping work and heat that maintain non-equilibrium steady states (NESS). Our results show that only an interaction in the energy-preserving form, or ``thermal operations'', lead to thermalization with the environment, while all other forms of interaction drive the system to a NESS. We then demonstrate that two classes of NESS are achievable: (i) those that settle to an effective temperature that is different from that of the environment. This effective temperature range extends to negative values, leading to non-passive steady-states with population inversion; (ii) those that exhibit steady-state coherences (SSC) in the energy eigenbasis. The latter arise due to a certain type of system-environment interaction that we fully characterize. For both cases, we examine the associated heat, work, and entropy currents. Our work aims at providing a unified and systematic study of the thermodynamics of a single system within the collision model framework for arbitrary interactions, thus complementing the current literature~\cite{StrasbergPRX}, which could in turn be generalized to arbitrary coupling strengths~\cite{SeahPRE2019}. Our work, in particular, reveals the possibility of creating {\it coherent} steady-states through system-ancilla interaction Hamiltonians, for which a thorough analysis of the corresponding thermodynamics, both in terms of cost (housekeeping heat and work) and of resources (extractable work), is provided.

The paper is organized as follows. We start by introducing memoryless collision models and recall the derivation of their associated master equation in Sec.~\ref{sec:cm}. In Sec.~\ref{sec:thermodynamics}, we lay out the preliminaries for characterizing the thermodynamics of open systems and establish that, generally, the system is driven into a NESS supported by non-zero work and heat currents and examine the exemplary case of a qubit collision model. blueSec.~\ref{SSCsection} focusses on the case of steady states that support coherences in the energy eigenbasis. There, we also characterize the ergotropy, which is the maximum amount of work one can extract from the NESS. Finally, we provide some concluding remarks in Sec.~\ref{sec:conclusion}.

\section{Memoryless Collision Models}
\label{sec:cm}
Let us consider a system $S$, with Hamiltonian $\hat{\Ham}_S$, coupled to an environment $E$ consisting of a series of identical non-interacting auxiliary units, each of them being described by the Hamiltonian $\hat{\Ham}_A$, so that $\hat{\Ham}_E\!\!=\!\!\sum_n \hat{\Ham}_A^{(n)}$. The free evolution of the system, generated by $\hat{\Ham}_S$, is punctuated by sequential interactions with the environmental units, each coupled to the system only once, for a time interval $\delta t$. This picture defines the class of repeated-interaction or collision models~\cite{Scarani2002, BuzekPRA, ziman2005b, ziman2010, ciccarello2017}. Although more general situations can be envisaged~\cite{StrasbergPRX, VacchiniPRL}, in what follows we will restrict to regular-in-time system-environment couplings and assume that the time intervals between subsequent collisions, during which the system evolves unitarily, are negligible compared to $\delta t$. This means that, after having interacted with the $n$-th environmental unit for a time interval $\delta t$, the system immediately moves on to interact with the $(n+1)$-th unit.

A precise formulation of the above scheme is given by taking the system-environment interaction to be externally changed in time according to
\begin{equation}
\label{eq:interactionHam}
\hat{\Ham}_{SE}(t)=\sum_{n} \Theta(t-n \delta t) \Theta\left((n+1) \delta t-t\right) \hat{\Ham}_{SA},
\end{equation}
where $\Theta(\cdot)$ denotes the Heaviside step function. A given collision changes the system according to
 \begin{equation}\label{eq:repeatevol}
\rho_{S}\left((n+1)\delta t\right)=\mathrm{Tr}_{A}\left[\hat{U}(\delta t) \,\left(\rho_{S}\left(n\delta t\right) \otimes \rho_{A}\right)\, \hat{U}^{\dagger}(\delta t)\right],
\end{equation}
where $\rho_A$ is the state of the $n$-th auxiliary unit and where the unitary operator is given by
\begin{equation}
\label{eq:unitary}
\hat{U}(\delta t) = e^{-i \delta t (\hat{\Ham}_S + \hat{\Ham}_A + \hat{\Ham}_{SA})} \, .
\end{equation}
Here and throughout, we chose units such that $\hbar\!=\!1$. As we will discuss in more detail in Section~\ref{sec:thermodynamics}, this picture allows us to neatly define and compute all the thermodynamic quantities such as heat, work, and entropy production, for arbitrary system-environment interactions. It is known that we can write down a master equation in GKSL form for the open system's evolution~\cite{LorenzoPRA, StrasbergPRX}. As such, one might naively expect that the corresponding thermodynamics would follow from the traditional formalism, see, e.g., Ref. \cite{Alicki, Spohn}, whereby all of the relevant quantities can be determined from the system state. However, as we discuss in the following, such a picture is only valid when the system-environment interaction is energy-preserving (i.e., it is in the rotating wave form), which is the only type of interaction where an equilibrium steady state (ESS) is achieved. Any other form of interaction gives rise to a NESS, which is maintained via steady state currents due to the energy invested in switching on and off the interaction in Eq.~\eqref{eq:interactionHam}.

\subsection{Continuous time limit}
\label{sec:GKSL}
Let us consider the most general interaction Hamiltonian of the form $\hat{\Ham}_{SA}\!=\!g \hat{V}_{SA}\!=\!g \sum_{ij} \hat{S}_i\otimes \hat{A}_j$, where $g$ is a common rescaling constant denoting the intensity of the coupling and where $\hat{S}_i (\hat{A}_j)$ are generic system (auxiliary unit) operators. Note that, in general, $\hat{V}_{SA}$ does not commute with $\hat{\Ham}_S $. Following ~\cite{LandiPRE, KarevskiPRL, LandiPRL}, it is convenient to assume that the coupling constant is proportional to the collision time according to $g\!\!\propto\!\! (\delta t)^{-1/2}$. We also assume that $\hat{\Ham}_S\!=\!J \sum_i \hat{S}_i $, where the system's characteristic frequency, $J$, sets the intrinsic system evolution timescale $t_S = J^{-1}$ over which the typical coherent dynamics generated by $\hat{\Ham}_S$ occurs. Furthermore, without any loss of generality~\cite{BreuerPetruccione, LorenzoPRA}, we assume that all the odd moments of the interaction Hamiltonian, $\hat{\Ham}_{SA}$, have zero average, i.e. $\mathrm{Tr}_{A}\left[\hat{\Ham}_{SA}^{2m+1}\rho_A\right]\!=\!0$. Finally, as stated above, we take $t_S \ll \delta t$, or, equivalently, $J\delta t \gg 1$; we are, thus, justified in neglecting the system's evolution between subsequent collisions. However, this does not mean that the system's free Hamiltonian is irrelevant, as it can play a crucial role during the interaction time~\cite{LorenzoPRA}.

We next briefly recall the derivation provided in Refs.~\cite{LorenzoPRA, StrasbergPRX}, where it has been shown that a closed effective master equation for the dynamics of the reduced system in the short collision time regime, $\delta t\!\ll\!1$, is obtainable. Moving to the interaction picture with respect to $\hat{\Ham}_A$, and expanding the unitary evolution operator Eq.~\eqref{eq:unitary} up to second order in $\delta t$, with $g=g_0 \delta t^{-1/2}$, leads to
\begin{equation}\label{eq:approx}
\hat{U}(\delta t) \simeq \Id_{SA} - i\left(\hat{\Ham}_S +\frac{g_0}{\sqrt{\delta t}}\hat{V}_{SA} \right) \delta t -g_0^2 \frac{\hat{V}_{SA}^{2}}{2\delta t} \delta t^{2},
\end{equation}
Substituting Eq.~\eqref{eq:approx} into Eq.~\eqref{eq:repeatevol} leads to the approximate expression for the change of the system's state due to a single collision
\begin{equation}
\begin{aligned}
\label{GKSL}
&\frac{\delta \rho_S \left(n\delta t\right)}{\delta t} \equiv \frac{\rho_S\left((n+1)\delta t\right) - \rho_S\left(n\delta t\right)}{\delta t} \notag\\
&~~~~~= -i\left[\hat{\Ham}_S,\rho_S\left(n\delta t\right)\right] + g_0^2 \mathrm{Tr}_{A}\left[\hat{V}_{SA}\,\left(\rho_S\left(n\delta t\right)\otimes\rho_A\right)\, \hat{V}_{SA}^{\dagger}\right] \\
&~~~~~- \frac{g_0^2}{2}\mathrm{Tr}_A\left[\left\{\hat{V}_{SA}^{\dagger}\hat{V}_{SA},\,\left(\rho_S\left(n\delta t\right)\otimes\rho_A\right)\right\}\right] \notag\\
&~~~~~= -i\left[\hat{\Ham}_S,\rho_S\left(n\delta t\right)\right] + \\
&~~~~~\sum_{jk} \gamma_{jk} \left( \hat{S}_j\rho_S\left(n\delta t\right)\hat{S}^{\dagger}_k- \frac{1}{2} \left\{\hat{S}^{\dagger}_k\hat{S}_j\,\rho_S\left(n\delta t\right)\right\}\right),
\end{aligned}
\end{equation}
where the explicit (but generic) form for $\hat{V}_{SA}$ introduced above has been used to obtain the last line, while $ \gamma_{jk}\!=\! g_0^2 \mathrm{Tr}_{A}\left[ \hat{A}^{\dagger}_k\hat{A}_j \right] $ is the damping rate determined by the environmental auto-correlation function~\cite{LorenzoPRA}. Taking the continuous time limit $\lim_{\delta t\to 0^+, n\to +\infty}\frac{\delta \rho_S\left(n\delta t\right)}{\delta t} \equiv \frac{d\rho_S(t)}{dt} $, with $t\!=\!n \delta t$ being finite, finally allows us to obtain the master equation
\begin{equation}\label{eq:GKSL}
\frac{d\rho_S(t)}{dt} = -i\left[\hat{\Ham}_S,\rho_S(t)\right] + \mathcal{L}(\rho_S(t)),
\end{equation}
where the dynamical generator is in GKSL form
\begin{equation}
\mathcal{L}(\rho_S(t)) \equiv \sum_{jk} \gamma_{jk} \left( \hat{S}_j\rho_S(t)\hat{S}^{\dagger}_k - \frac{1}{2} \left\{\hat{S}^{\dagger}_k\hat{S}_j\,\rho_S(t)\right\}\right) \, .
\end{equation}
Note that the convenient rescaling we have chosen for the interaction $\hat{\Ham}_{SA}\!\!=\!\!g \hat{V}_{SA}$ with $g\!\propto\!(\delta t)^{-1/2}$ implies that the GKSL master equation, Eq.~\eqref{eq:GKSL}, has been derived in the ultra-strong coupling regime. In other words, since it is valid for $\delta t\!\to\!0^+$, it requires a diverging coupling strength to ensure a meaningful contribution from the system-environment interaction. Such an approach in deriving a GKSL master equation, alongside the familiar weak coupling Born-Markov-secular approximation~\cite{BreuerPetruccione}, is reminiscent of the singular coupling limit~\cite{AlickiLendiBook}, where the environmental correlation function is $\delta$-correlated in time.

The structure of Eq.~\eqref{eq:GKSL} for the generator of the reduced dynamics has several important consequences. It guarantees that the dynamical map $\rho_S(0)\!\mapsto\!\rho_S(t)\!\equiv\! \Lambda(t,0)[\rho_S(0)]$, with $\Lambda(t,0)\!=\!e^{t\mathcal{L}}$, is completely positive and trace preserving, thus representing a physical operation. Furthermore, it implies semigroup composition law $\Lambda(t,s)\Lambda(s,0)\!=\!\Lambda(t,0)$. This, in turn, characterizes the dynamics as Markovian, i.e. memoryless, according to all of the criteria introduced in the literature~\cite{VacchiniRMP}. Additionally, the set of steady-states are constructed as the kernel of the dynamical generator, i.e. the set of operators $\rho^*$ such that $\mathcal{L}(\rho^*)\!=\!0$. This set always consists of at least one element, due to the trace-preserving nature of the dynamical map. However, a Lindblad master equation can have multiple steady-states (e.g. in the pure-dephasing dynamics). Whenever the steady-state $\rho^*$ is non-unique, then some information about the initial state is preserved in the infinite-time limit. We refer to Refs.~\cite{albert2014symmetries, denisov2019universal}, and references therein, for further details and properties of semigroup dynamical maps and Lindblad generators. The main focus of the present work is, rather, to characterize the thermodynamic properties of steady-states in the special case where this form of Lindblad dynamics stems from a collision model.

There are manifestly two categories of steady-states: equilibrium steady states (ESS) and non-equilibrium steady states (NESS). From the thermodynamics point of view, a principle difference between the two is the presence of non-zero average entropy production for NESS. This implies that these states support non-zero currents for thermodynamic quantities. In order to establish a natural framework to discuss the thermodynamics of the system, we consider each auxiliary unit that makes up the environment to start in a thermal state  $\rho_A\!\equiv\!\rho^\text{th}_A\!= \!Z_A^{-1} e^{-\beta\hat{\Ham}_A}$ with $Z_A\!=\!\mathrm{Tr}\left[e^{-\beta\hat{\Ham}_A}\right]$. Such states are a subset of KMS states that can be shown to satisfy the KMS condition put forward to define thermal equilibrium~\cite{arXiv_Alicki,BreuerPetruccione}. In the present context of open quantum systems, the KMS condition translates into the following relation on the environmental decay function $\gamma_{jk}(-\omega) \!=\! e^{-\beta\omega} \gamma_{kj}(\omega)$. This condition, when combined with a GKSL master equation \textit{derived under the secular approximation}, which decouples the evolution of the populations from that of the coherences by averaging over the fast oscillating terms in the master equation, leads to detailed balance. As a result, the stationary state of the dynamics is the thermal state $\rho^* \equiv \rho^\text{th}_S\!=\!Z_S^{-1} e^{-\beta\hat{\Ham}_S}$ and thermal equilibration with the bath is achieved, leaving the system in an ESS. 

Crucially however, in our case the master equation Eq.~\eqref{eq:GKSL} in GKSL form was obtained {\it without} performing the secular approximation, owing to the rescaling of the system-environment interaction Hamiltonian and the fast-collision time limit $\delta t \to 0$. Therefore, the equations of motion for the populations and the coherences are not necessarily decoupled from each other, and this means that even if the state of the environment satisfies the KMS condition, detailed balance is not fulfilled and, therefore, the stationary state is given by some form of NESS. In the following Sections, we characterize the types of achievable NESS, showing that we can even reach stationary states with non-zero values of coherences in the system's energy eigenbasis. For the class of collision models considered here, an ESS is reached only when the system-environment interaction is energy preserving, which is the only choice of system-environment interaction for which the aforementioned equations of motion do indeed decouple, so that one ends up with $\rho^* \!=\! \rho^\text{th}_S$, while all other interactions lead to a NESS. As a consequence, we establish that one must carefully account for the switching on and off of the interaction term in order to accurately capture the thermodynamics of the process. Such a work cost has also been shown to be crucial in the consistency of the thermodynamic analysis of coupled quantum systems~\cite{GabrieleNJP2018}. 

\section{Thermodynamics of Open Quantum Systems}
\label{sec:thermodynamics}
Consider the typical setting of a system with Hamiltonian $\hat{\Ham}_S(t)$, interacting with an environment with Hamiltonian $\hat{\Ham}_E$ via a coupling term $\hat{\Ham}_{SE}(t)$. The time-dependence in $\hat{\Ham}_S(t)$ and $\hat{\Ham}_{SE}(t)$ is assumed to arise due to some external control protocol, and therefore acts only at the level of the system, which usually incorporates the controllable degrees of freedom. The overall dynamics is governed by the unitary operator
\begin{equation}
 U(t) = \overrightarrow{\mathrm{T}} \exp\left[ -i \int_0^t d\tau \hat{\Ham}_{tot}(t) \right],
\end{equation}
with $\hat{\Ham}_{tot} (t)\!\!=\!\!\hat{\Ham}_S(t) + \hat{\Ham}_E + \hat{\Ham}_{SE}(t) $. Assume, further, that system and environment are initially uncorrelated, i.e. $\rho_{SE}(0)\!\!=\!\!\rho_S(0) \otimes \rho_E(0)$, and, in order to meaningfully study the thermodynamic properties, that the environment is a thermal bath with a generic inverse temperature $\beta\!\in\![-\infty, +\infty]$,
\begin{equation}
\rho_E(0) = \rho^\text{th}_E \equiv Z_E^{-1} e^{-\beta\hat{\Ham}_E}~\text{with}~~Z_E = \mathrm{Tr}_E\left[e^{-\beta\hat{\Ham}_E}\right].
\end{equation}
The work performed by the external agent responsible for the explicit time dependence of the Hamiltonian can be defined as the change in the total energy of the composite system, induced by the action of the external protocol
\begin{align}
\label{workdef}
 W(t) &= \mathrm{Tr}_{SE}\left[\hat{\Ham}_{tot} (t) \rho_{SE}(t)\right] - \mathrm{Tr}_{SE}\left[\hat{\Ham}_{tot} (0) \rho_{SE}(0)\right] \notag\\
 &= \int_0^t d\tau \frac{d}{d\tau}\left(\mathrm{Tr}_{SE}\left[ \hat{\Ham}_{tot}(\tau) \rho_{SE}(\tau)\right]\right)\notag\\
 &= \int_0^t d\tau \mathrm{Tr}_{SE}\left[ \frac{d\hat{\Ham}_{tot}(\tau) }{d\tau}\rho_{SE}(\tau)\right]\notag\\
 &= \int_0^t d\tau \mathrm{Tr}_{SE}\left[ \left(\frac{d\hat{\Ham}_{S}(\tau) }{d\tau}+\frac{d\Ham_{SE}(\tau)}{d\tau}\right)\rho_{SE}(\tau)\right]
\end{align}
where we have used
\begin{equation}\label{conservation}
\mathrm{Tr}_{SE}\left[ \hat{\Ham}_{tot}(\tau)\frac{d\rho_{SE}(\tau)}{d\tau} \right] = 0
\end{equation}
since $\frac{d\rho_{SE}(\tau)}{d\tau}\!=\! -i\left[\hat{\Ham}_{tot}(\tau),\rho_{SE}(\tau)\right]$ and due to the assumption that the Hamiltonian of the environment is time independent. The heat can be analogously defined as the change in the energy of the bath~\cite{ReebWolfNJP, EspositoNJP2010},
\begin{align}\label{heatdef}
 Q(t) &= \mathrm{Tr}_E\left[\hat{\Ham}_E(\rho_E(t) - \rho_E(0))\right] \notag\\
 &= \int_0^t d\tau \mathrm{Tr}_{E}\left[ \hat{\Ham}_E \frac{d\rho_E(\tau)}{d\tau}\right].
\end{align}
As the overall system is considered closed, its evolution is unitary and the first law of thermodynamics is satisfied. From Eqs.~\eqref{workdef} and \eqref{heatdef} the change in the internal energy is
\begin{align}\label{intendef}
\Delta E(t) &= \mathrm{Tr}_{SE}\left[(\hat{\Ham}_{S} (t) + \hat{\Ham}_{SE} (t)) \rho_{SE}(t)\right]  \notag\\
&\qquad \qquad-  \mathrm{Tr}_{SE}\left[(\hat{\Ham}_{S} (0) + \hat{\Ham}_{SE} (0)) \rho_{SE}(0)\right]\notag\\
&= \int_0^t d\tau  \frac{d}{d\tau}\left(\mathrm{Tr}_{SE}\left[(\hat{\Ham}_{S} (\tau) + \hat{\Ham}_{SE} (\tau)) \rho_{SE}(\tau)\right]\right),
\end{align}
which is well-defined independently of the strength of the coupling between system and bath. At the level of the second law, one finds that the entropy production is given by
\begin{equation}
\label{entrproddef}
\Sigma(t) = \beta Q(t) - \Delta S(t) = D\left(\rho_{SE}(t) || \rho_S(t)\otimes\rho^\text{th}_E\right),
\end{equation}
where $ \Delta S(t)\!=\!S(\rho_S(0)) - S(\rho_S(t))$ is the change in the system's von Neumann entropy $S(\rho)\!=\!-\mathrm{Tr}\left[ \rho\ln\rho \right]$, and where $D(\rho||\sigma)\!=\! \mathrm{Tr}\left[\rho\ln\rho\right] - \mathrm{Tr}\left[\rho\ln\sigma\right]$ is the relative entropy. In Refs.~\cite{ReebWolfNJP, EspositoNJP2010} it has been shown that the irreversible entropy production can be equivalently calculated as
\begin{equation}
\Sigma(t) = I\left(\rho_S(t), \rho_E(t)\right) + D\left(\rho_E(t) || \rho^\text{th}_E\right),
\end{equation}
where $I\left(\rho_S(t), \rho_E(t)\right)\!\equiv\! S(\rho_S(t)) + S(\rho_E(t)) - S(\rho_{SE}(t))$ is the mutual information between the system and the environment (see also Refs.~\cite{PaternostroNPJ, GooldPRE}). This expression shows that two terms contribute to the entropy production, one is correlation between $S$ and $E$ built up during the coupled evolution and the other is the change in the environmental state. The application of Klein's inequality~\cite{BreuerPetruccione} immediately leads to the conclusion that the entropy production is a positive quantity. Its time derivative however, i.e. the entropy production rate, is not generally constrained to be positive. Transient non-Markovian dynamics is known to lead to negative entropy production rates~\cite{LeggioPRE, StevePRA, MarcantoniSciRep, PatiPRA} although a strict relationship between these notions is still lacking~\cite{StrasbergPRE, FengPRE}.

\subsection{Weak coupling limit}
Taking one extreme of the singular coupling limit, i.e., the oft-considered weak coupling limit, where the interaction Hamiltonian $\hat{\Ham}_{SE}$ is proportional to a coupling constant that is much smaller than any energy scale of $\hat{\Ham}_S$ and $\hat{\Ham}_E$, we find that the work, Eq.~\eqref{workdef}, reduces to
\begin{align}
\label{workWC}
 W(t) &\simeq \int_0^t d\tau \mathrm{Tr}_{SE}\left[ \left(\frac{d\hat{\Ham}_{S}(\tau) }{d\tau}\right)\rho_{SE}(\tau)\right] \notag\\
 &= \int_0^t d\tau \mathrm{Tr}_{S}\left[ \left(\frac{d\hat{\Ham}_{S}(\tau) }{d\tau}\right)\rho_{S}(\tau)\right]
\end{align}
and the change in the internal energy is
\begin{align}
\label{intenWC}
\Delta E(t) &\simeq \int_0^t d\tau \frac{d}{d\tau}\left(\mathrm{Tr}_{S}\left[ \hat{\Ham}_{S}(\tau) \rho_{S}(\tau)\right]\right)
\end{align}
From the first law it follows that the heat is given by
\begin{align}
\label{heatWC}
Q(t) & \simeq  \int_0^t d\tau \mathrm{Tr}_{S}\left[ \hat{\Ham}_{S}(\tau) \frac{d}{d\tau}\left(\rho_{S}(\tau)\right)\right],
\end{align}
all of which is consistent with the standard definitions put forward in Ref.~\cite{Alicki}. Note that the last expression could be equivalently derived from Eq.~\eqref{heatdef} by writing $\hat{\Ham}_E\!=\! \hat{\Ham}_{tot} - \hat{\Ham}_S - \hat{\Ham}_{SE}$, neglecting the last term due to the weak coupling approximation and using Eq.~\eqref{conservation}. This implies that, in the weak-coupling limit, all of the thermodynamic quantities can be calculated directly from the reduced system only, once the generator of the dynamics is known. Indeed, the entropy production, in this limit, can be written as $\Sigma (t)\!=\!\int_0^t d\tau \sigma(\tau)$, with the rate $\sigma$ given by
\begin{align}
\sigma(t) &= \frac{d}{dt}\left(\Delta S(t) - \beta Q(t)\right) \notag\\
&= -\mathrm{Tr}_S\left[ \frac{d}{dt}\rho_S(t) \left(\ln\rho_S(t) + \beta\hat{\Ham}_S\right) \right].
\end{align}
Using the thermal state of $S$, $\rho^\text{th}_S$, we can rewrite the above expression as
\begin{equation}\label{entprodrateWC}
\sigma(t) = \frac{d}{dt}D\left(\rho_S(t) || \rho^\text{th}_S\right).
\end{equation}
Notice that $\rho^\text{th}_S$ is the stationary state of the dynamics, i.e. $\frac{d\rho^\text{th}_S}{dt}\!=\!0$, when the Born-Markov and secular approximations are assumed to be valid. As the weak coupling limit results in a master equation in GKSL form, Spohn's inequality~\cite{Spohn} guarantees that $\sigma(t)\!\geq\!0$ $\forall t$.

We stress that Eqs.~\eqref{workWC}, \eqref{intenWC}, \eqref{heatWC} and \eqref{entprodrateWC} are valid only in the weak coupling regime. In what follows, we show that despite the dynamics being described by a master equation with the same general form, these expressions cannot be applied {\it verbatim} to describe the thermodynamics of collision models considered in this work.

\subsection{Thermodynamics of collision models}
Returning to the collision models described by Eq.~\eqref{eq:repeatevol}, we remark that while we do not consider any explicit time dependence in the system's Hamiltonian, the full system-environment interaction is in fact time dependent by virtue of Eq.~\eqref{eq:interactionHam}. We will find that this is the sole source of work that maintains generic NESS. We assume all environmental units to be initially prepared in the Gibbs thermal state $\rho^\text{th}_A$. Evaluating Eq.~\eqref{workdef} over a single collision gives us an expression for the energetic cost which can be written in terms of the work required to switch on and off the interaction Hamiltonian
\begin{equation}
\label{workcollision}
W(\delta t) = \mathrm{Tr}_{SA}\left[ \left( \hat{U}^{\dagger}(\delta t)\hat{H}_{SA} \hat{U}(\delta t) - \hat{H}_{SA} \right) \rho_S(0)\otimes\rho^\text{th}_A \right].
\end{equation}
This term can only vanish when the interaction commutes with the free Hamiltonian, which occurs only for the energy-preserving case typically considered in the literature~\cite{BarraPRE2017}. Analogously, the heat dissipated into an environmental unit during a single collision is
\begin{equation}
\label{heatcollision}
Q(\delta t) = \mathrm{Tr}_{SA}\left[ \left( \hat{U}^{\dagger}(\delta t)\hat{H}_{A} \hat{U}(\delta t) - \hat{H}_{A} \right) \rho_S(0)\otimes\rho^\text{th}_A \right].
\end{equation}
The change in the internal energy after each collision as well as the entropy production follow immediately, where in particular $\Sigma (\delta t)\!=\! \Delta S(\delta t) - \beta Q(\delta t)$, with $\Delta S(\delta t)\!\equiv\! S(\rho_S(\delta t)) - S(\rho_S(0))$. Notice that we have made use of the fact that for the memoryless collision model with fixed $\delta t$ these quantities do not depend on the specific collision, $n$.

It is interesting and instructive to consider how Eqs.~\eqref{workcollision} and \eqref{heatcollision} change in the continuous time limit $n\!\!\to\!\!\infty$ and $\delta t\!\to\! 0^+$ such that $t \!\equiv\!n\delta t$ is finite, which corresponds to the ultra-strong coupling limit, where the evolution of the reduced system is also governed by a master equation in GKSL form, Eq.~\eqref{GKSL}. In this limit, Eqs.~\eqref{workcollision} and ~\eqref{heatcollision} can be simplified by using Eq.~\eqref{eq:approx}, leading to $W(t)\!\!=\!\!\int_0^t d\tau \dot{W}(\tau)$ and $Q(t)\!=\!\int_0^t d\tau \dot{Q}(\tau)$ with
\begin{widetext}
\begin{align}
&\dot{W}(t) \equiv \lim_{\delta t\to 0^+} \frac{W(\delta t)}{\delta t} = \mathrm{Tr}_{SA}\left[ \left( \hat{\Ham}_{SA}\left( \hat{\Ham}_S +\hat{\Ham}_A\right) \hat{\Ham}_{SA} - \frac{1}{2}\left\{  \hat{\Ham}^2_{SA},\left( \hat{\Ham}_S +\hat{\Ham}_A\right)\right\}\right)\rho_S(t)\otimes\rho^\text{th}_A\right],\\
&\dot{Q}(t) \equiv \lim_{\delta t\to 0^+} \frac{Q(\delta t)}{\delta t} = \mathrm{Tr}_{SA}\left[ \left( \hat{\Ham}_{SA}\hat{\Ham}_A \hat{\Ham}_{SA} - \frac{1}{2}\left\{  \hat{\Ham}^2_{SA},\hat{\Ham}_A\right\}\right)\rho_S(t)\otimes\rho^\text{th}_A\right].
\end{align}
\end{widetext}
Finally, when the system has reached the steady-state the changes in both the internal energy $\Delta U(t)$ and the entropy $\Delta S(t)$ will vanish, while work and heat will become equal and opposite. In the cases where the interaction is energy preserving, the steady-state $\rho^*$ will be the thermal state $\rho^\text{th}_S$, and, consequently, both heat and work will also vanish. This is the only situation where thermal equilibrium is reached. For any other interaction Hamiltonian that does not preserve the bare energy $\hat{\Ham}_S + \hat{\Ham}_A$, a NESS is attained, which supports finite heat and work currents. Once integrated, these quantities provide an elegant example of the \textit{housekeeping heat and work} ~\cite{OonoPaniconi, SasaTasaki, DeffnerPRE}, which are necessary to maintain the steady-state out of equilibrium. These quantities, if calculated over an infinite time interval, diverge linearly as a result of the fact that the currents become constants in the steady-state. In the following, we will elucidate these results for paradigmatic interactions.

\begin{figure*}[t]
\includegraphics[width=0.5\columnwidth]{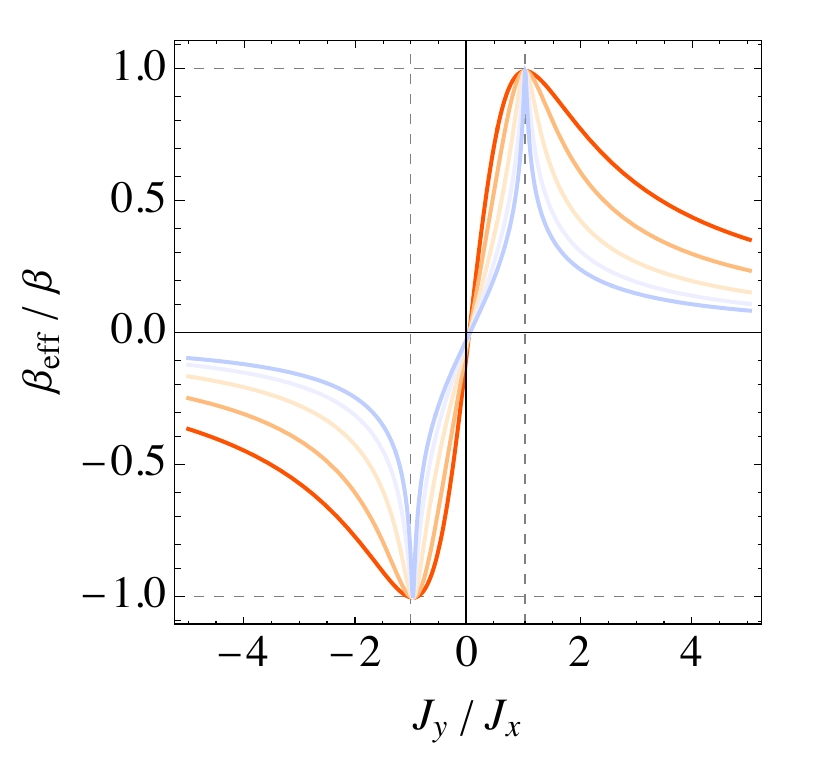}
\includegraphics[width=0.52\columnwidth]{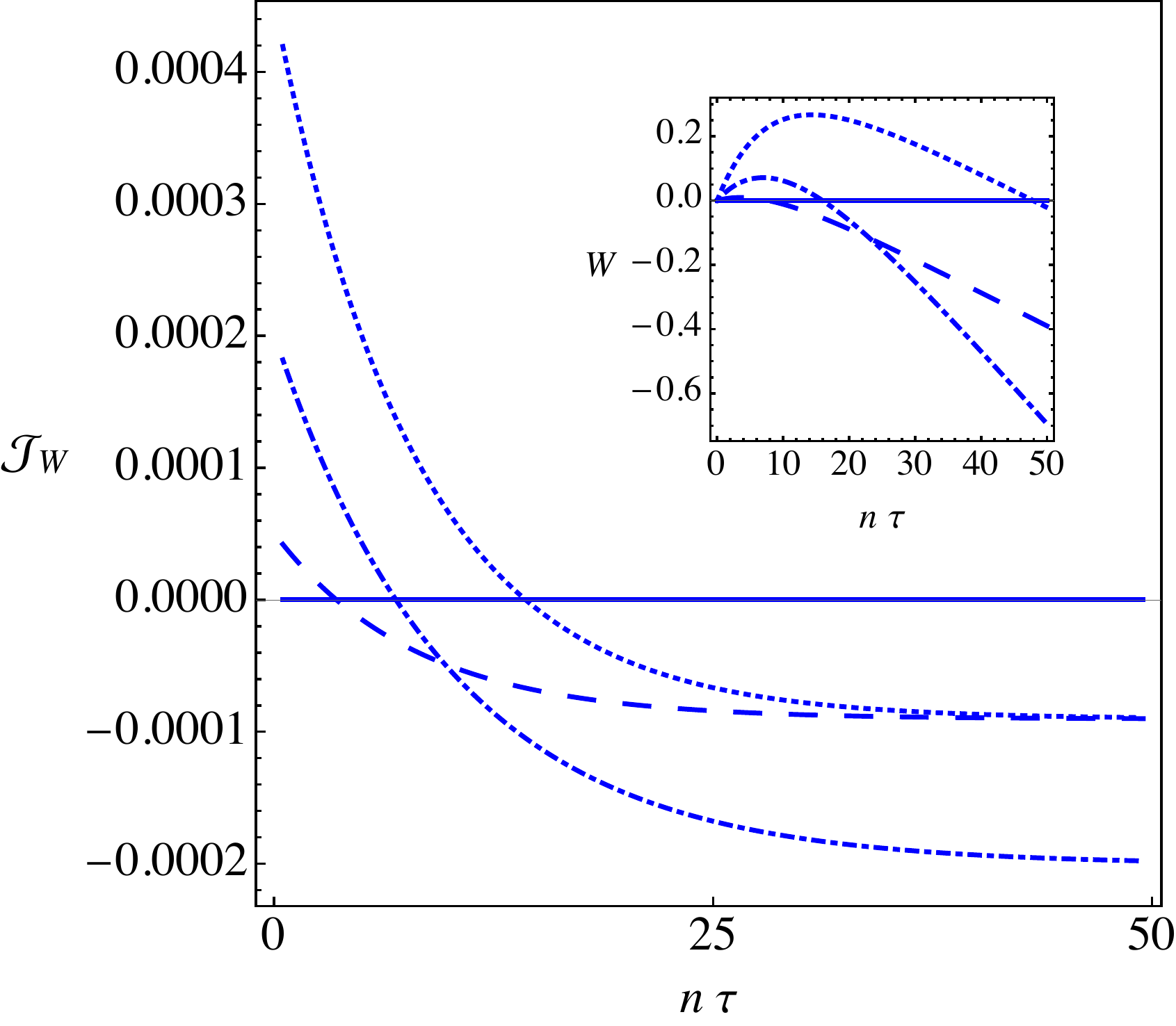}
\includegraphics[width=0.52\columnwidth]{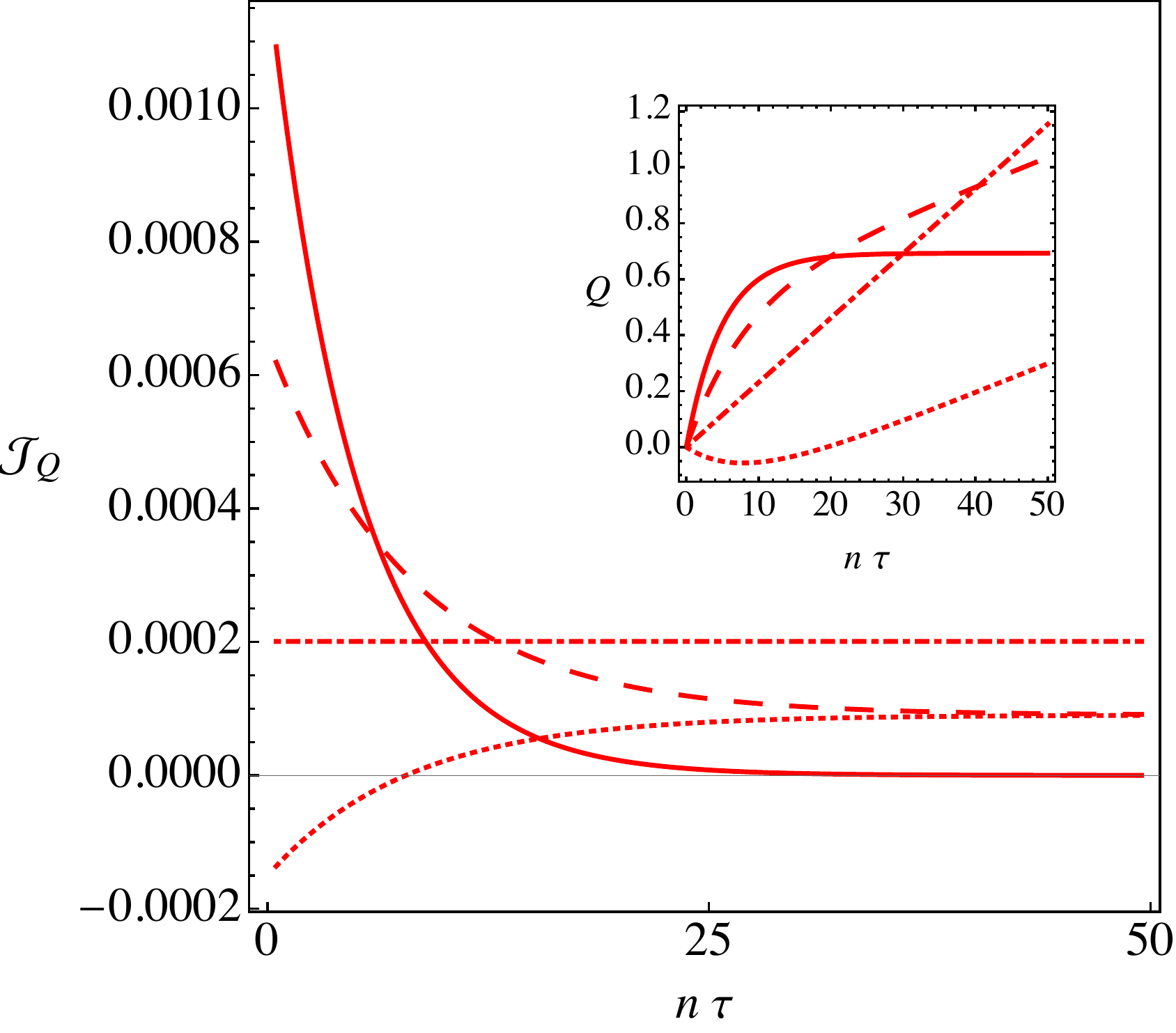}
\includegraphics[width=0.48\columnwidth]{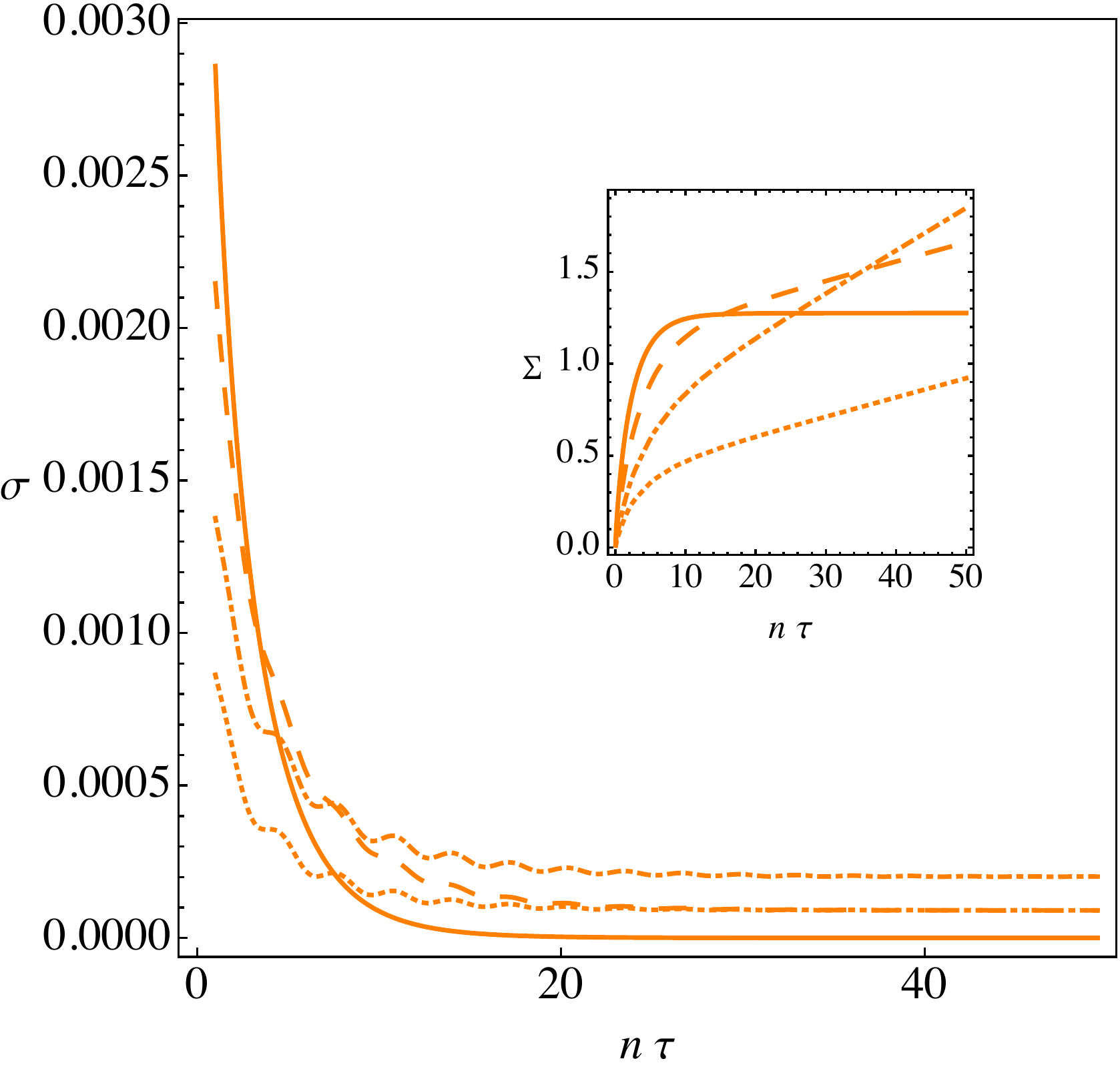}
\caption{(a) Ratio of the effective steady state temperature with the environment temperature as a function of the ratio of the coupling strengths, $J_y/J_x$. The color gradient from lighter-blue to darker-red denotes the increasing temperature of the environment going from $\beta\!=\!9$ to $\beta\!=\!1$ in steps of 2. (b) The work, (c) heat and (d) entropy currents with their corresponding actual (integrated) values as the insets. In each panel we fix $\beta = 1$ and the ratio between coupling strengths to be $J_y/J_x\!=\!-1/2$ [dotted], 0 [dot-dashed], 1/2 [dashed], and 1 [solid]. We have (arbitrarily) fixed $\rho_S(0) = \lvert \Psi(\theta) \rangle\langle \Psi(\theta)\rvert$, with $\lvert \Psi(\theta) \rangle = \cos(15\pi/16) \lvert 1 \rangle + \sin(15\pi/16)\lvert 0\rangle$. While the transient behaviors will be affected for different initial states, the steady state behavior shown remains the same.}
\label{Fig3}
\end{figure*}

\subsection{Qubit Collision Model: Diagonal steady states in the energy eigenbasis}
\label{sec:results}
Let us now demonstrate the above framework in the exemplary setting of a qubit collision model. In what follows, we recapitulate and expand on some known results in the literature~\cite{StrasbergPRX, BarraPRE2017, SeahPRE2019, Onat2019}. We will consider the system and all auxiliary units to be two-level quantum systems, with their respective free Hamiltonians given by $\hat{\Ham}_{S(A)}\!\!=\!\!\frac{\omega_{S(A)}}{2}\hat{\sigma}_z$, where $\lbrace \hat{\sigma}_x, \hat{\sigma}_y, \hat{\sigma}_z \rbrace$ are the usual Pauli matrices. We chose the generic two-body interaction
\begin{equation}
\hat{\Ham}_{SA} = \sum_{lm} J_{lm} \hat{\sigma}_l\otimes\hat{\sigma}_m.
\end{equation}
As our focus is on steady state properties we will assume an arbitrarily chosen $\rho_S(0)$ (see Sec.~\ref{SSCsection} for discussions on initial state dependence), nevertheless the qualitative behavior exhibited holds for any  suitable choice of initial state. The environmental qubits are prepared in the Gibbs state $\rho_A(0)\!=\!\rho^\text{th}_A $ with
\begin{equation}
\rho^\text{th}_A = \begin{pmatrix}
\frac{1}{2}\left[1-\tanh\left(\frac{\beta\omega_A}{2}\right)\right] & 0 \\ 0 & \frac{1}{2}\left[1+\tanh\left(\frac{\beta\omega_A}{2}\right)\right]
\end{pmatrix}
\end{equation}
and $\rho_{SE}(0)\!=\!\rho_S(0) \bigotimes \rho^\text{th}_A$.
In the following numerical results, the continuous time limit is achieved by taking the collision time $\delta t\!\!\sim\!\!5* 10^{-2} \omega_S$ and $n\!=\!10^3$, such that $t\!\!=\!\!n \delta t\!\!=\!\!50\omega_S$. This guarantees that we are well within the steady-state regime dictated by the master equation.

Let us start by considering the diagonal interaction case, i.e.
\begin{equation}
\label{diagonalInteraction}
\hat{\Ham}_{SA} = J_x \hat{\sigma}_x\otimes\hat{\sigma}_x + J_y \hat{\sigma}_y\otimes\hat{\sigma}_y,
\end{equation}
where the term $\propto\!\hat{\sigma}_z\otimes\hat{\sigma}_z$ has been neglected as it essentially amounts to a simple phase factor. When $J_x\!=\!J_y$, and the system and auxiliary qubits are resonant, i.e. $\omega_S\!=\!\omega_A$, the interaction is energy preserving, and, therefore, we have vanishing heat and work currents at the steady-state
\begin{equation}
\dot{W}(t) = \dot{Q}(t) = 0.
\end{equation}
The resulting stationary state is the equilibrium Gibbs distribution at the same inverse temperature $\beta$ of the environment, meaning that we achieved an ESS. This is the only interaction for which Eqs.~\eqref{workWC}, \eqref{intenWC}, \eqref{heatWC} and \eqref{entprodrateWC} accurately capture the thermodynamics of the process and constitutes the type of interaction typically considered in the literature~\cite{RuariPRA, BarraSciRep, StrunzPRA2016, PezzuttoNJP, BarraPRE2017, BarisPRA2019, CampbellNJP2019}.

For any other setting governed by Eq.~\eqref{diagonalInteraction} the system will reach a NESS, $\rho^*$, which is diagonal in the energy eigenbasis. As our system is a qubit, it is possible to express the steady-state as a Gibbs state at an effective inverse temperature $\beta_{\text{eff}}$ such that
\begin{equation}
\rho^* = \rho^\text{th}_S(\beta_{\text{eff}}) = \frac{e^{-\beta_{\text{eff}}\hat{\Ham}_S}}{\mathrm{Tr}_S\left[e^{-\beta_{\text{eff}}\hat{\Ham}_S}\right]}.
\end{equation}
In particular, for a balanced but off-resonance scenario, i.e. $J_x\!=\!J_y$ but $\omega_S\!\neq\!\omega_A$, $\beta_{\text{eff}}$ can be found simply through the relation $\beta_{\text{eff}} \omega_S\!=\!\beta\omega_A$. Therefore, this situation essentially amounts to a re-normalization of the effective temperature.

A more interesting situation occurs when the system and the environmental units are resonant, but $J_x\!\neq\!J_y$. In this case, the effective temperature has a more involved dependence as a function of the ratio $ J_y/J_x$. In Fig.~\ref{Fig3}(a) the various curves correspond to different bath temperatures, with the gradient from lighter-blue to darker-red corresponding to increasing temperature. For $J_y/J_x\!=\!0$ the steady state has an infinite effective temperature, $\beta_{\text{eff}}\!=\!0$, meaning that the stationary state $\rho^*$ is the maximally mixed state. In the region $0<J_y/J_x\!<\!1$, $\beta_\text{eff}$ increases with the ratio of the coupling strengths $J_y/J_x$, eventually reaching thermal equilibrium only when $ J_y/J_x\!=\!1$. Further increasing the ratio, $J_y/J_x\!>\!1$, the steady-state temperature increases, and it asymptotically approaches to infinity as $J_y/J_x\rightarrow\infty$. Notice that $|\frac{\beta_{\text{eff}}}{\beta}|\!\leq\!1$ for all the possible values of coupling strengths $J_y/J_x$. This indicates that the system is always driven to a higher effective temperature than the bath. We also see that cold environments maintain NESS that are further away from equilibrium, relative to the bath temperature. Interestingly, when $J_x$ and $J_y$ have different signs, i.e. $J_y/J_x\!<\!0$, the system settles to a NESS with inverted populations, indicated by a negative $\beta_{\text{eff}}$. This behavior is present for all initial environment temperatures and, importantly, does not require the auxiliary qubits to be inverted, it is solely controlled by $\hat{\Ham}_{SE}$. The crucial difference between these NESS and those that attain a positive temperature is that the former are active states~\cite{Pusz}. Therefore, work can be extracted from them via unitary cyclic processes~\cite{AlickiPRE, AllahverdyanEPL, BinderNJP, CampbellBatteries}.

\begin{figure*}[t]
\includegraphics[width=0.47\columnwidth]{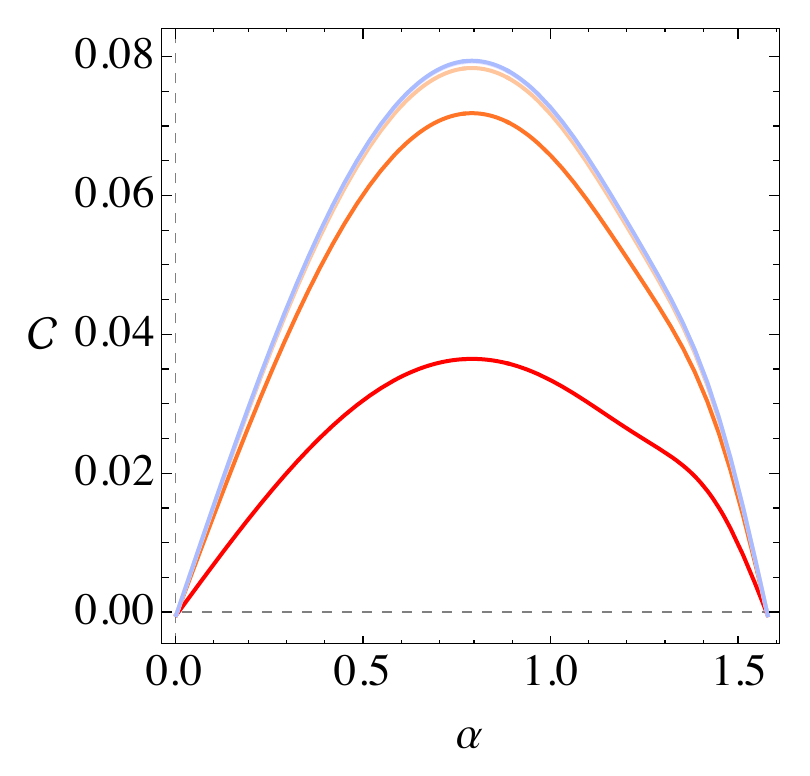}
\includegraphics[width=0.53\columnwidth]{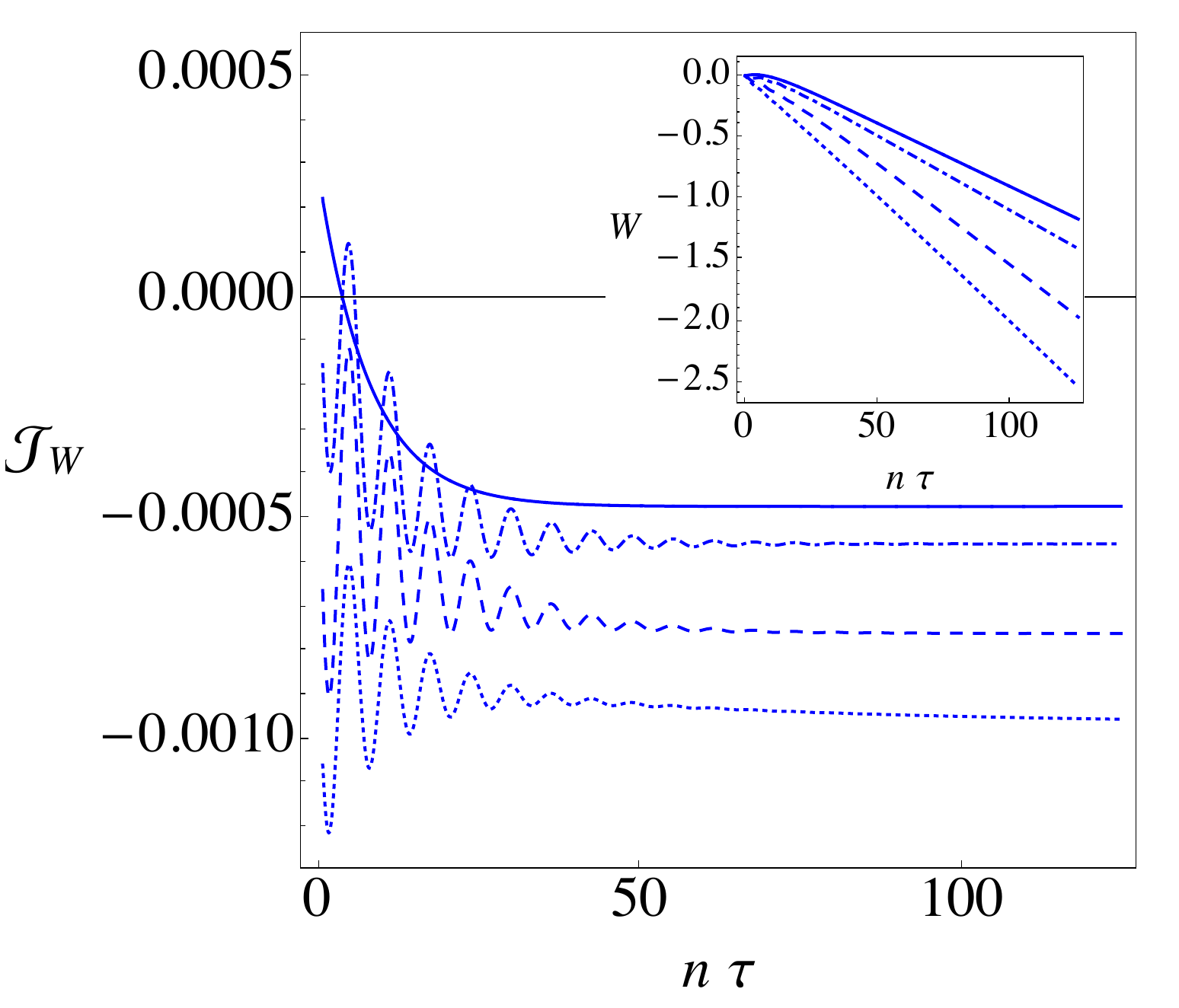}
\includegraphics[width=0.51\columnwidth]{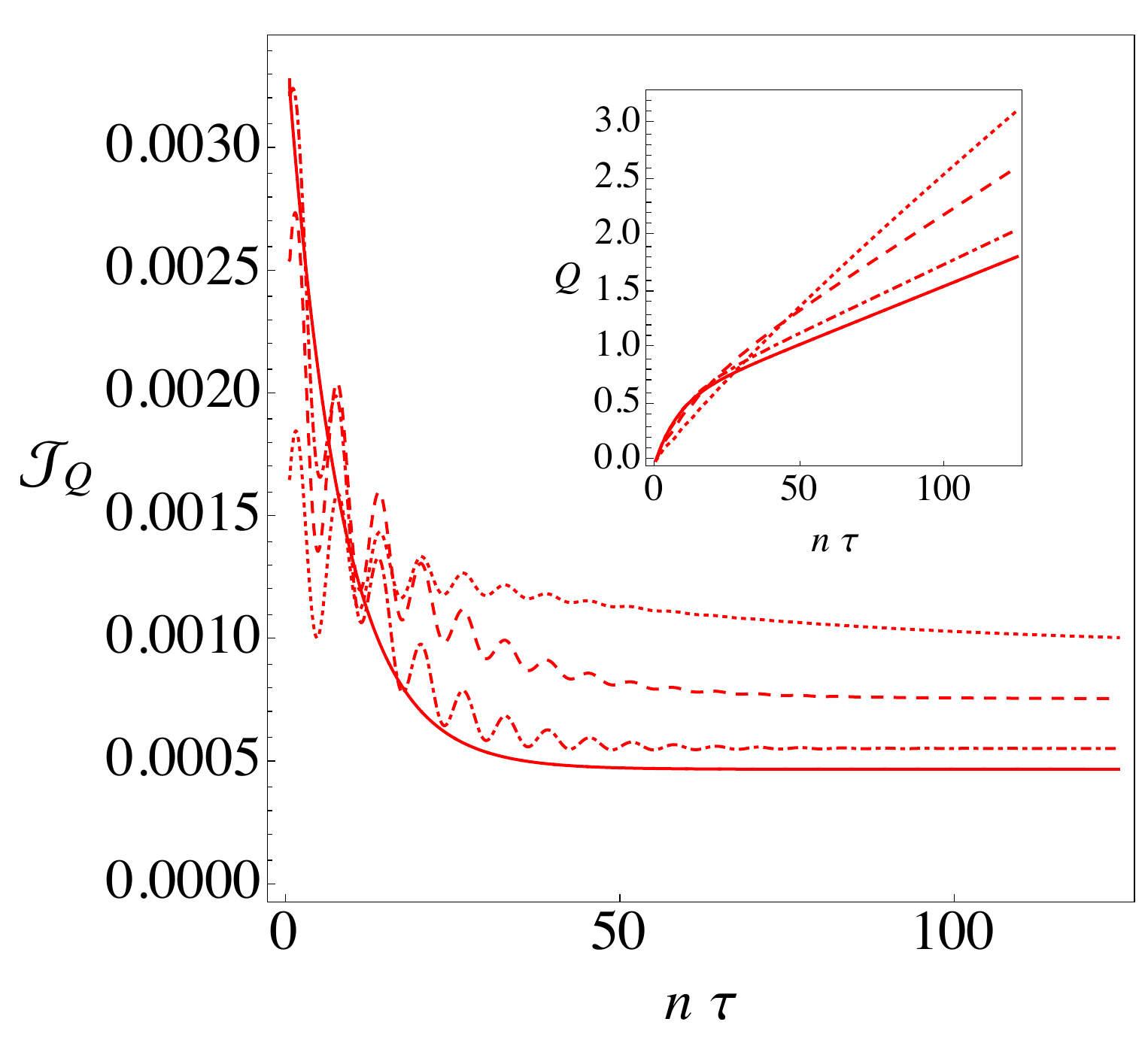}
\includegraphics[width=0.48\columnwidth]{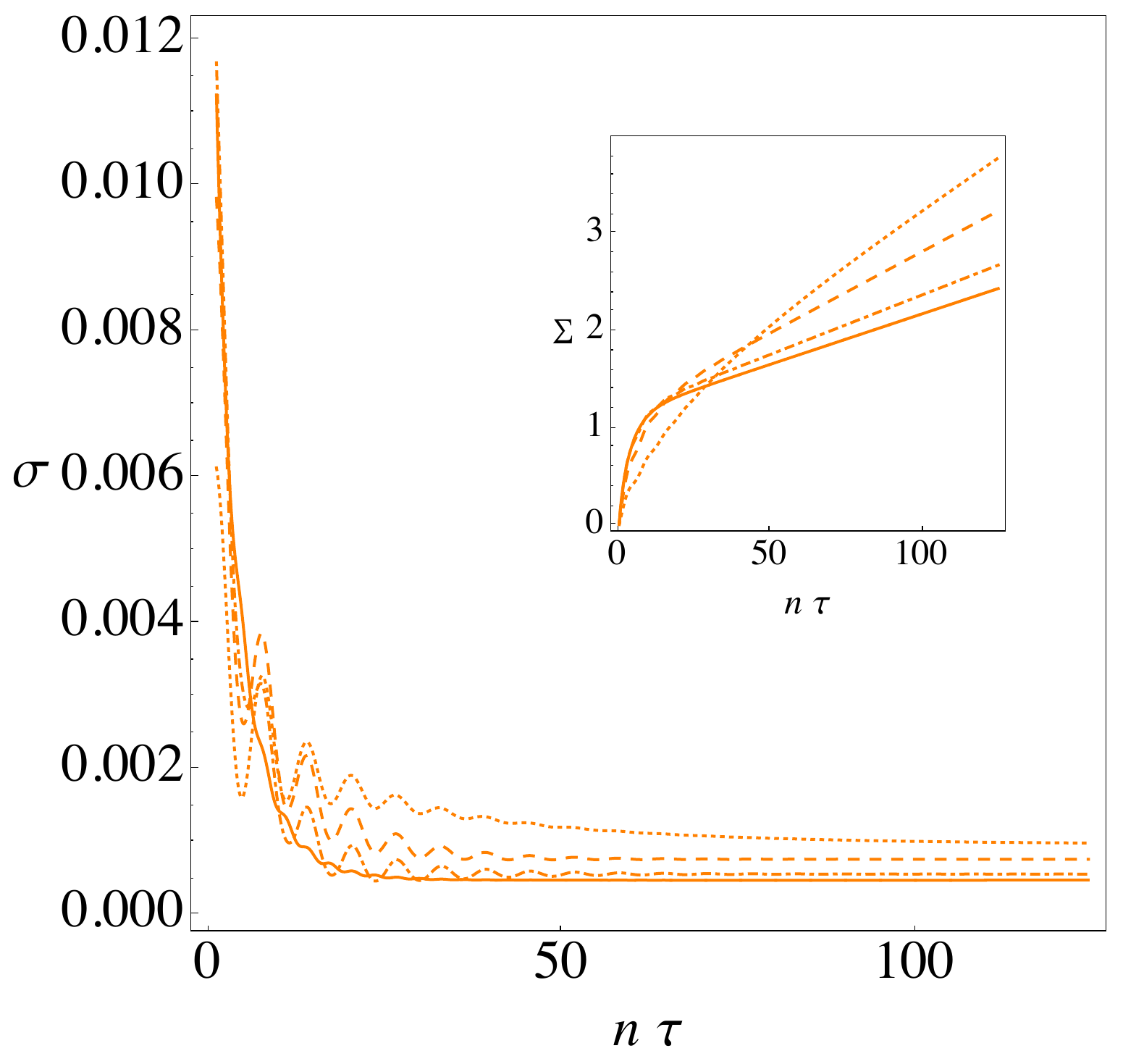}
\caption{(a) $l_1$-norm of coherence as a function of $\alpha$ for various environment temperatures going from $\beta\!=\!9$ (top-most lighter blue) to $\beta\!=\!1$ (bottom-most darker-red) in steps of 2. (b) The work, (c) heat and (d) entropy currents with their corresponding integrated quantities shown as insets with $\beta\!=\!1$. In each panel, we fix $\alpha\!=\!0$ [solid], $\pi/8$ [dotted], $\pi/4$ [dashed], and $3\pi/8$ [dot-dashed]. The initial state is as for Fig.~\ref{Fig3}. In all panels we have fixed $J_x\!=\!2J_y\!=\!1$.}
\label{Fig5}
\end{figure*}

Turning our attention to the thermodynamic quantities, through Eqs.~\eqref{workcollision} and \eqref{heatcollision}, we show the work, heat, and entropy production in Fig.~\ref{Fig3} as insets of their respective currents, as a function of time $t\!=\!n\delta t$, for different choices of $J_y/J_x$ and for resonant system and environment $\omega_S\!=\!\omega_A$. When the system-environment interaction preserves the total energy, i.e. $J_y\!=\!J_x$ (solid curves), the energetic cost of switching the interaction on and off vanishes as evidenced in Fig.~\ref{Fig3}(b), where both the work and its current remain zero throughout. During the transient, panel (c) shows that a finite amount of heat enters the system, thermalizing it to the environment temperature, and, therefore, bringing the system to the ESS. The entropy production, panel (d), behaves consistently with these findings, and, crucially, we observe that both heat and entropy currents vanish in the steady-state regime, as the system has reached the ESS. This is consistent with the fact that, in this case, system and environment return to a fully factorized state~\cite{GiovannettiPRA2018}.

Shifting our focus to the unbalanced interactions, we find the underlying mechanisms supporting the NESS. From Fig.~\ref{Fig3}, it is clear that for such interactions the steady-state heat, work and entropy production have constant, non-zero currents. Therefore, all of these quantities keep changing linearly with time. The currents that remain in the steady-state are clear examples of housekeeping heat and work that keep the system out of equilibrium~\cite{OonoPaniconi, SasaTasaki, DeffnerPRE}. Since the system-environment interaction does not preserve the total energy, it is the work associated to switching the interaction on and off between consecutive collisions that is the source for supporting the NESS. We find that higher effective temperatures for the system imply larger heat, work and entropy currents, with the maximum occurring for $J_y/J_x\!=\!0$ (dot-dashed). Thus, the magnitude of housekeeping currents increases when the system is kept further out of equilibrium with respect to the bath.

\section{Coherent non-equilibrium steady states}
\label{SSCsection}

Next, we examine the possibility to generate steady-states with coherence in the energy eigenbasis. Our interest is two-fold: first, the interaction of a quantum system with its environment generally results in the loss of its quantum properties; therefore introducing a way of achieving coherence in an open system setting is interesting in its own right. Second, from the thermodynamic point of view, similarly to the NESS at a negative temperature discussed previously, steady-states with coherence are also non-passive and work can be extracted from them~\cite{KorzekwaNJP}. Therefore, we also examine the ergotropy of the full range of available steady states and, for the special class of interactions that generate coherences, show that an initial state dependence can emerge when the interaction becomes strongly dephasing.

\subsection{Steady State Features}
Owing to the generality of the system-environment interaction in the derivation of our master equation, we find that generating steady-state coherences (SSC) is possible for interaction Hamiltonians of the form
\begin{equation}\label{SSC}
\hat{\Ham}_{SA} = (J_x \hat{\sigma}_x\otimes\hat{\sigma}_x + J_y \hat{\sigma}_y\otimes\hat{\sigma}_y ) + J_{zy} \hat{\sigma}_z\otimes\hat{\sigma}_y.
\end{equation}
Notice that this Hamiltonian belongs to the class of interactions put forward in Ref.~\cite{GuarnieriPRL}, as it consists of both a parallel term with respect to $\hat{\Ham}_S$, $\hat{\sigma}_z\otimes\hat{\sigma}_y$, and a perpendicular one, $J_x\hat{\sigma}_x\otimes\hat{\sigma}_x + J_y \hat{\sigma}_y\otimes\hat{\sigma}_y$. We stress that `parallel' and `perpendicular' are intended with respect to the Hilbert-Schmidt scalar product on the space of system operators. As we established in the previous section, the perpendicular term will drive the system to a diagonal state in the energy eigenbasis, while the parallel term leads to dephasing. Either term on their own cannot support SSC; however, when the two are combined, the competition between them drives the system generally to a NESS with SSC. It is also worth remarking that, while the perpendicular term leads any initial state to a unique steady state, this is no longer strictly true for Eq.~\eqref{SSC}. This is due to the well established fact that for pure dephasing, the steady state has a strong dependence on the initial state. Thus, the introduction of the parallel term in Eq.~\eqref{SSC} introduces an initial state dependence. However, this effect is small for $J_{zy}\!\!<\!\!\sqrt{J_x^2 + J_y^2}$, i.e. when the perpendicular term is the dominant interaction. In order to quantify the amount of coherences, we will employ the $l_1-$ norm of coherence $\mathcal{C}\!=\!\sum_{l\neq m} |\rho^*_{lm}|$, first introduced in \cite{Baumgratz2014}, which can be shown to satisfy all the properties to be considered as a valid coherence measure~\cite{Adesso2017RMP}. Fig.~\ref{Fig5}(a) shows its dependence on $\alpha\!=\!\arctan\left(\frac{J_{yz}}{\sqrt{J_x^2+J_y^2}}\right)$, for different environment temperatures. When either of the two couplings goes to zero, the SSC vanish, cf. Fig.~\ref{Fig5}(a). Furthermore, in agreement with the results obtained in~\cite{GuarnieriPRL, ArchakarXiv}, the maximum amount of SSC is obtained when $\alpha\!=\! \pi/4$ and for zero temperature environments. The amount of SSC monotonically decreases with increasing bath temperature.

Turning to Fig.~\ref{Fig5}(b)-(d), we see that for the chosen interaction term, Eq.~\eqref{SSC}, the NESS is supported by non-zero equal and opposite heat and work currents for any value of $\alpha$. Furthermore, the asymptotic value of the heat current and of the entropy production rate $\sigma$ are increased by the presence of SSC with respect to the reference value (solid line) for $\alpha=0$. This increment represents the precise energetic cost that is paid in order to generate and maintain the SSC. It is furthermore worth noticing that the switching on of the additional parallel term $\propto\! \hat{\sigma}_z\otimes\hat{\sigma}_y$ induces transient oscillations in the thermodynamic currents, which is reflected in a more marked difference in the integrated quantities, cf. the insets. Finally, a special case is obtained when $J_y = 0$ in Eq.~\eqref{SSC}. In this regime, the steady-state values for all the above currents of heat, work and entropy production are not affected by the building up of SSC. Although the transient behavior still shows oscillations, the asymptotic value for the currents is purely determined in this case by $J_x\hat{\sigma}_x\otimes\hat{\sigma}_x$, i.e. the dot-dashed curves in Fig.~\ref{Fig3}.

To summarize, from this analysis we find that for the special class of interaction Hamiltonians outlined in Ref.~\cite{GuarnieriPRL} a system can be driven to a NESS with SSCs. The magnitude of these coherences is dependent on a delicate interplay between the parallel and perpendicular components of the system-ancilla interaction Hamiltonian. The maximal amount of coherence is achievable when the weights of both such terms are equal to each other. The thermodynamic consequences of the generation and maintenance of SSC in collision models is then reflected in the currents of housekeeping heat, work and entropy production.

\subsection{Ergotropy}
The previous sections have shown that the nature of the system environment interaction term generally leads to various classes of NESS which are supported by non-zero work and heat currents. Moreover, we have quantified the thermodynamic cost of driving the system to a target NESS by computing the associated heat, work and entropy production. As discussed above, such quantities represent the housekeeping costs, i.e. the amount of energy one needs to invest to create and maintain non-equilibrium configurations. On a similar but complementary note, one can now ask how much work can be extracted from these states in a scenario when the repeated interaction scheme is stopped once the system has reached the steady-state regime and the NESS is a non-passive state.

The key figure of merit which quantifies the maximum amount of work that can be gained from a quantum system by means of cyclic unitary transformations is known as \textit{ergotropy} ~\cite{ErgotropyEPL}. Given a generic Hamiltonian $\Ham_S \!=\! \sum_k \epsilon_k \ket{\epsilon_k}\bra{\epsilon_k}$, and a non-passive state $\rho_S = \sum_j r_j \ket{r_j}\bra{r_j}$, this quantity is given by
\begin{equation}
\mathcal{E} = \sum_{jk} r_j\epsilon_k \left( |\langle {r_j}\!\ket{\epsilon_k}|^2 - \delta_{jk} \right).
\end{equation}
To examine the full range of achievable states, we parameterize the Hamiltonian, Eq.~\eqref{SSC} to be
\begin{equation}
\begin{aligned}
&\hat{\Ham}_{SA} =\cos(\alpha) (\cos(\gamma) \hat{\sigma}_x\otimes\hat{\sigma}_x + \sin(\gamma) \hat{\sigma}_y\otimes\hat{\sigma}_y ) +\sin(\alpha) \hat{\sigma}_z\otimes\hat{\sigma}_y, \\
&\text{with}~\alpha\!=\!\arctan\left(\frac{J_{yz}}{\sqrt{J_x^2+J_y^2}}\right)~~\text{and}~~\gamma = \arctan\left(J_y / J_x\right)
\end{aligned}
\end{equation}
As discussed in the previous section, varying $\alpha$ and $\gamma$ allows to reach NESS with population inversion and/or with SSC.

\begin{figure}[t]
{\bf (a)} \hskip0.45\columnwidth {\bf (b)}
\includegraphics[width=0.55\columnwidth]{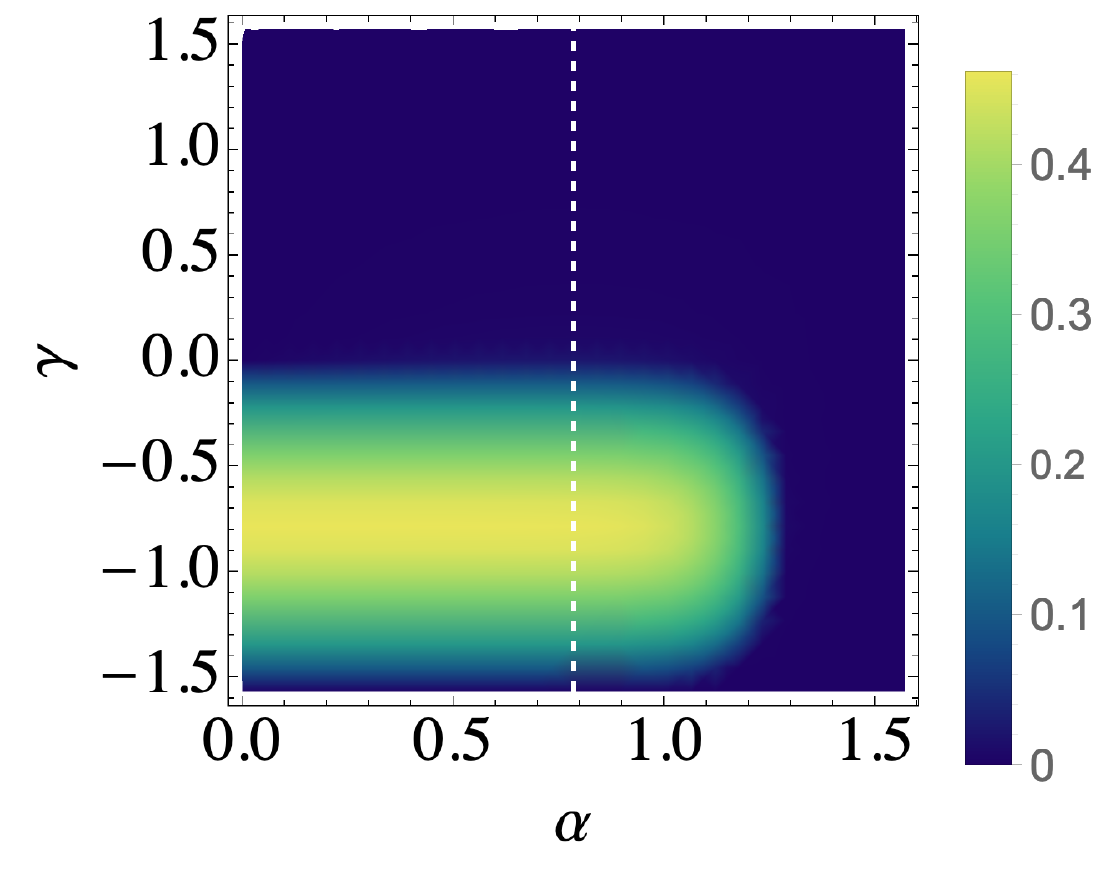}~\includegraphics[width=0.45\columnwidth]{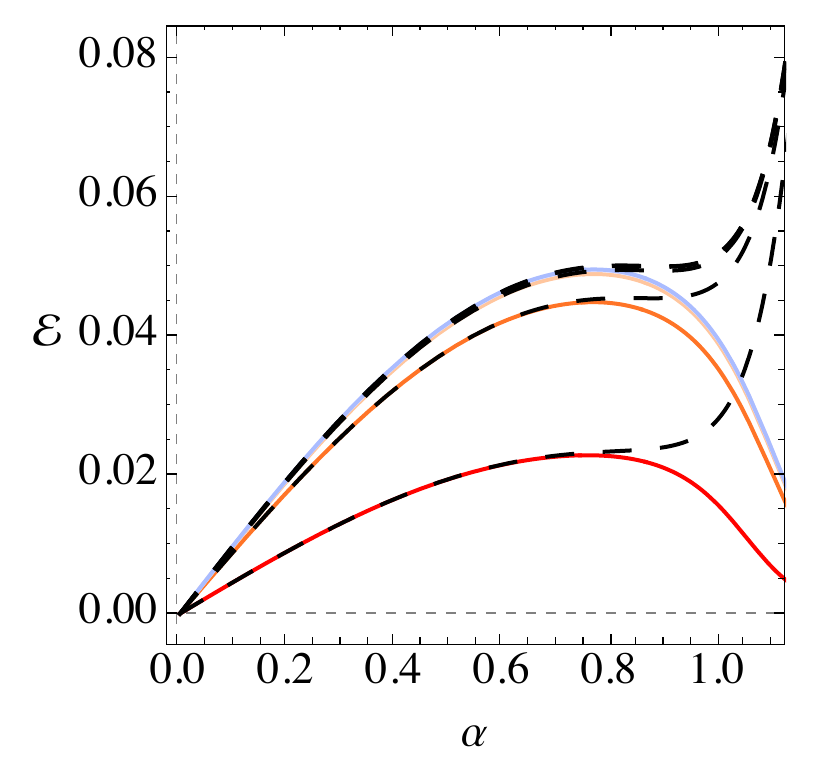}
\caption{(a) Ergotropy of the steady-state as a function of interaction strengths. The $x$-axis accounts for the introduction of the coherence generating term $J_{zy}$ while the $y$-axis corresponds to varying the strength of the $J_y$ term. The initial state is fixed to be $|0\rangle$. $\gamma\!=\!-\pi/4$ and $\alpha\!=\!0$ corresponds to the state with highest inverted populations. The region to the right of the vertical dashed line is when the coupling is dominated by the $J_{zy}$ term and the ergotropy becomes strongly dependent on the choice of initial state. (b) We fix $\gamma\!=\!0$ and show the extremal values of the ergotropy for several values of $\beta\!=\!1$ [bottommost]  to 9 [topmost] in steps of 2. The solid (dashed) curves correspond to $\rho_S(0)\!=\!\ket{0}\!\bra{0}$ $\left(\rho_S(0)\!=\!\ket{1}\!\bra{1}\right)$. Note that we truncate the range of $\alpha$ for visibility.}
\label{figergotropy}
\end{figure}

Fig.~\ref{figergotropy}(a) shows the full behavior of ergotropy as a function of $\alpha$ and $\gamma$ for fixed inverse temperature $\beta\!=\!1$. We clearly see that the largest ergotropy corresponds to population inverted steady-states. Indeed for $\alpha\!<\!\pi/4$, as the magnitude of the achievable SSC is small, we find they do not noticeably contribute to the ergotropy. For $\alpha\!>\!\pi/4$, we find that the ergotropy becomes significantly dependent on the initial state of the system. This is due to the fact that in this regime, $\mathcal{H}_{SA}$ is dominated by the parallel term; therefore, the open dynamics of the system is closer to that of pure dephasing. In Fig.~\ref{figergotropy}(b), we fix $\gamma\!=\!0$ and show the extremal values of $\mathcal{E}$ for the system initialized in $\ket{0}$ (solid curves) and $\ket{1}$ (dashed curves). We clearly see that for $\alpha\!<\!\pi/4$ the extractable work is largely independent of the initial state. This is due to the fact that, in this regime, the steady state is principally dictated by the perpendicular term, which tends to drive all initial states to the same steady state. For our chosen parameter values, this means that the populations are almost equal and, consequently, the extractable work in this regime is completely due to the presence of the SSC. However, when $\alpha\!>\!\pi/4$, since the populations are less affected by the dynamics,  initial states with large populations in their excited states tend to persist and, therefore, we can see a strong dependence of ergotropy on the initial state in this regime. It is finally worth pointing out that if one starts from a generic passive initial state, the energetic cost of creating and maintaining a given NESS is always (and, typically, much) higher than the amount of extractable work as quantified by the ergotropy, in accordance with the second law.

\section{Conclusions}
\label{sec:conclusion}

We have examined in detail the thermodynamics of memoryless quantum collision models through a master equation description derived in the ultra-strong coupling regime and for any system-environment interaction. By carefully computing the associated energy exchanges at play, we showed that, despite the dynamics being fully described by a master equation in GKSL form, the corresponding thermodynamics cannot be described by properties of the system alone, in contrast to the weak-coupling limit. The root of this discrepancy lies in the fact that the secular approximation is not taken and, as such, the equations of motion for populations and coherences do not decouple, with the notable exception of energy-preserving interactions, which is the only instance where the system reaches thermal equilibrium with the bath.

We have demonstrated our results for an all-qubit version of our collision model, explicitly showing that both equilibrium and non-equilibrium steady-states (NESS) are admitted by the dynamics. We have shown the rich variety of NESS that emerge, broadly falling into two distinct categories: those that are diagonal in the energy eigenbasis and those that exhibit steady-state coherences (SSC). Among the NESS achievable, those with negative effective temperatures and those with SSC particularly stand out. The former implies that it is possible to achieve a population inversion, while the latter is one of the few examples of coherence generation through an interaction with an environment. These states are examples of non-passive states and therefore one can extract a non-zero amount of work from them. We established that all NESS are supported by non-zero work, heat, and entropy currents. These quantities account for the cost of maintaining steady states out-of-equilibrium with the environment, and, therefore, provide an elegant demonstration of the housekeeping work and heat. Our work therefore contributes to the ongoing effort to understand the thermodynamics of strongly coupled quantum systems~\cite{Dou2018, Rivas2019, BenentiPRA}.

\acknowledgements
G. G., B.\c{C}. and S. C. gratefully acknowledge discussions with Kant, Adalet, Belk\i s, and Qubit. We are grateful to Gabriele De Chiara, John Goold, Gabriel Landi, Mark Mitchison, Stefan Nimmrichter, Philip Strasberg and Bassano Vacchini for useful discussions. G. G. acknowledges the European Research Council Starting Grant ODYSSEY (Grant Agreement No. 758403).  D. M. acknowledges the Erasmus Programme for support. B.\c{C}. acknowledges support from The Research Fund of Bah\c{c}e\c{s}ehir University (BAUBAP) Project No: BAP.2019.02.03. S. C. gratefully acknowledges the Science Foundation Ireland Starting Investigator Research Grant “SpeedDemon” (No. 18/SIRG/5508) for financial support.

\bibliography{collisions_SSC}
\end{document}